\documentclass[aip,graphicx]{revtex4-1}
\usepackage [latin1]{inputenc}
\draft 

\begin{document}

\title{New symmetry for the imperfect fluid} 

\author{Alcides Garat}
\affiliation{1. Former Professor at Universidad de la Rep\'{u}blica, Av. 18 de Julio 1824-1850, 11200 Montevideo, Uruguay.}
\email[]{garat.alcides@gmail.com}
\date{\today}

\begin{abstract}
\begin{center}
{\bf Abstract}
\end{center}
We will address the existence of a new symmetry for an imperfect fluid by introducing local four-velocity gauge-like transformations for the case when there is vorticity. A similar tetrad formulation as to the Einstein-Maxwell spacetimes formalism presented in previous manuscripts will be developed in this manuscript for the imperfect fluids. The four-velocity curl and the metric tensor will be invariant under these kind of four-velocity gauge-like local transformations. While the Einstein-Maxwell stress-energy tensor is locally gauge invariant under electromagnetic gauge transformations, the perfect fluid stress-energy tensor will not be invariant under four-velocity gauge-like local transformations. We will dedicate our analysis to the imperfect fluid stress-energy tensor that will be invariant under local four-velocity gauge-like transformations when additional transformations are introduced for several variables included in the stress-energy tensor itself. We will also pay special attention to the construction of a vorticity stress-energy tensor invariant under local four-velocity gauge-like transformations. An application on neutron stars will be developed in order to show the simplifications brought about by these new tetrads.
\end{abstract}


\maketitle 

\section{Introduction}
\label{intro}

We aim to study the imperfect fluid four-dimensional Lorentzian curved spacetimes with signature (-+++).
However in our present case we will start our analysis with an ideal fluid where the stress-energy tensor can be described by the following equation \cite{EG,AC,F},

\begin{equation}
T_{\mu\nu}= (\rho + p)\:u_{\mu}\:u_{\nu} + p\:g_{\mu\nu}\ ,\label{SET}
\end{equation}

where $\rho$ is the energy-density of the fluid, $p$ the isotropic pressure and $u^{\mu}$ its four-velocity field, $g_{\mu\nu}$ is the metric tensor. If in addition this fluid has vorticity $\omega_{\mu\nu}$, then we can proceed to build the new tetrads for this particular case following the method developed in reference \cite{A} and explained from different perspectives in references \cite{IWCP,AEO,ACD,ENV,MW}. These new tetrads will manifestly and covariantly diagonalize the stress-energy tensor (\ref{SET}) at every spacetime event.

We introduce the fluid extremal field or the velocity curl-extremal field through the local duality transformation given by,

\begin{equation}
\xi_{\mu\nu} = \cos\alpha \:\: u_{[\mu;\nu]} - \sin\alpha \:\: \ast u_{[\mu;\nu]} ,\label{vef}
\end{equation}

where $\ast u_{[\mu;\nu]} = {1 \over 2}\:\epsilon_{\mu\nu\sigma\tau}\:g^{\sigma\rho}\:g^{\tau\lambda}\:u_{[\rho;\lambda]}$ is the dual tensor of $u_{[\mu;\nu]}$ and the local complexion $\alpha$ is defined through the condition

\begin{equation}
\xi_{\mu\nu}\:\ast \xi^{\mu\nu} = 0 \ .\label{fcond}
\end{equation}

The symbol $;$ stands for covariant derivative with respect to the metric tensor $g_{\mu\nu}$. The identity,

\begin{eqnarray}
A_{\mu\alpha}\:B^{\nu\alpha} -
\ast B_{\mu\alpha}\: \ast A^{\nu\alpha} &=& \frac{1}{2}
\: \delta_{\mu}^{\:\:\:\nu}\: A_{\alpha\beta}\:B^{\alpha\beta}  \ .\label{i1}
\end{eqnarray}

which is valid for every pair of antisymmetric tensors in a four-dimensional Lorentzian spacetime \cite{MW}, when applied to the case $A_{\mu\alpha} = \xi_{\mu\alpha}$ and $B^{\nu\alpha} = \ast \xi^{\nu\alpha}$ yields an equivalent condition to (\ref{fcond}),

\begin{equation}
\xi_{\mu\rho}\:\ast\xi^{\mu\lambda} = 0 \ .\label{scond}
\end{equation}

The complexion, which is a local scalar, can then be found by plugging equation (\ref{vef}) in equation (\ref{fcond}) to be,

\begin{equation}
\tan(2\alpha) = - \left( u_{[\mu;\nu]}\:g^{\sigma\mu}\:g^{\tau\nu}\:\ast u_{[\sigma;\tau]}\right) \:/\: \left(u_{[\lambda;\rho]}\:g^{\lambda\alpha}\:g^{\rho\beta}\: u_{[\alpha;\beta]}\right) \ .\label{complexion}
\end{equation}

After introducing the new velocity curl-extremal field we proceed to write the four orthogonal vectors that will become an intermediate step in constructing the tetrad that diagonalizes the stress-energy tensor (\ref{SET}),

\begin{eqnarray}
V_{(1)}^{\alpha} &=& \xi^{\alpha\lambda}\:\xi_{\rho\lambda}\:X^{\rho}
\label{V1}\\
V_{(2)}^{\alpha} &=& \xi^{\alpha\lambda} \: X_{\lambda}
\label{V2}\\
V_{(3)}^{\alpha} &=& \ast \xi^{\alpha\lambda} \: Y_{\lambda}
\label{V3}\\
V_{(4)}^{\alpha} &=& \ast \xi^{\alpha\lambda}\: \ast \xi_{\rho\lambda}
\:Y^{\rho}\ ,\label{V4}
\end{eqnarray}

In order to prove the orthogonality of the tetrad (\ref{V1}-\ref{V4}) it is necessary to use the identity (\ref{i1}) for the case $A_{\mu\alpha} = \xi_{\mu\alpha}$ and $B^{\nu\alpha} = \xi^{\nu\alpha}$, that is,

\begin{eqnarray}
\xi_{\mu\alpha}\:\xi^{\nu\alpha} - \ast \xi_{\mu\alpha}\: \ast \xi^{\nu\alpha} &=& \frac{1}{2}
\: \delta_{\mu}^{\:\:\:\nu}\:Q\ ,\label{i2}
\end{eqnarray}

where $Q=\xi_{\mu\nu}\:\xi^{\mu\nu}$ is assumed not to be zero. We are free to choose the vector fields $X^{\alpha}$ and $Y^{\alpha}$, as
long as the four vector fields (\ref{V1}-\ref{V4}) do not become trivial. Let us introduce some names. The tetrad vectors have two essential components. For instance in vector $V_{(1)}^{\alpha}$ there are two main structures. First, the skeleton, in this case $\xi^{\alpha\lambda}\:\xi_{\rho\lambda}$, and second, the gauge vector $X^{\rho}$. In vector $V_{(3)}^{\alpha}$ the skeleton is $\ast \xi^{\alpha\lambda}$, and the gauge vector $Y_{\lambda}$. It is clear that if our choice for these fields is $X^{\alpha} = Y^{\alpha} = u^{\alpha}$, then the following orthogonality relations will hold,

\begin{eqnarray}
\lefteqn{ g_{\rho\mu}\:u^{\rho}\:V_{(2)}^{\mu} = g_{\rho\mu}\:u^{\rho}\:\xi^{\mu\lambda}\:u_{\lambda} = 0 \label{ortho1} } \\
&&g_{\rho\mu}\:u^{\rho}\:V_{(3)}^{\mu} = g_{\rho\mu}\:u^{\rho}\:\ast\xi^{\mu\lambda}\:u_{\lambda} = 0 \ ,\label{ortho2}
\end{eqnarray}

because of the antisymmetry of the velocity curl-extremal field $\xi_{\mu\nu}$. Then, at the points in spacetime where the set of four vectors (\ref{V1}-\ref{V4}) is not trivial, we can proceed to normalize,

\begin{eqnarray}
\hat{U}^{\alpha} &=& \xi^{\alpha\lambda}\:\xi_{\rho\lambda}\:u^{\rho} \:
/ \: (\: \sqrt{-Q/2} \: \sqrt{u_{\mu} \ \xi^{\mu\sigma} \
\xi_{\nu\sigma} \ u^{\nu}}\:) \label{Uw}\\
\hat{V}^{\alpha} &=& \xi^{\alpha\lambda}\:u_{\lambda} \:
/ \: (\:\sqrt{u_{\mu} \ \xi^{\mu\sigma} \
\xi_{\nu\sigma} \ u^{\nu}}\:) \label{Vw}\\
\hat{Z}^{\alpha} &=& \ast \xi^{\alpha\lambda} \:  u_{\lambda} \:
/ \: (\:\sqrt{u_{\mu}  \ast \xi^{\mu\sigma}
\ast \xi_{\nu\sigma}   u^{\nu}}\:)
\label{Zw}\\
\hat{W}^{\alpha} &=& \ast \xi^{\alpha\lambda}\: \ast \xi_{\rho\lambda}
\: u^{\rho} \: / \: (\:\sqrt{-Q/2} \: \sqrt{ u_{\mu}
\ast \xi^{\mu\sigma} \ast \xi_{\nu\sigma}  u^{\nu}}\:) \ .
\label{Ww}
\end{eqnarray}

In analogy with the electromagnetic case and without altering anything fundamental we assume for simplicity that $u_{\mu} \ \xi^{\mu\sigma} \ \xi_{\nu\sigma} \ u^{\nu} > 0$, $\:\:\:u_{\mu}
\ast \xi^{\mu\sigma} \ast \xi_{\nu\sigma}  u^{\nu} > 0$ and $-Q > 0$. We also assume that $\hat{U}^{\alpha}\:\hat{U}_{\alpha}=-1$. In terms of these tetrad vectors (\ref{Uw}-\ref{Ww}) and applying the method developed in manuscript \cite{A} we can express the four-velocity curl in its maximal simple form,

\begin{equation}
u_{[\mu;\nu]} = -2\:\sqrt{-Q/2}\:\:\cos\alpha\:\:\hat{U}_{[\alpha}\:\hat{V}_{\beta]} +
2\:\sqrt{-Q/2}\:\:\sin\alpha\:\:\hat{Z}_{[\alpha}\:\hat{W}_{\beta]}\ .\label{VC}
\end{equation}

The metric tensor will be written as,

\begin{equation}
g_{\alpha\beta} = -\hat{U}_{\alpha}\:\hat{U}_{\beta} + \hat{V}_{\alpha}\:\hat{V}_{\beta} +
\hat{Z}_{\alpha}\:\hat{Z}_{\beta} + \hat{W}_{\alpha}\:\hat{W}_{\beta}\ .\label{MT}
\end{equation}

The pair of vectors ($\hat{U}^{\alpha}, \hat{V}^{\alpha}$) span the local plane one. The vectors ($\hat{Z}^{\alpha}, \hat{W}^{\alpha}$) span the local orthogonal plane two. When we introduce gauge-like transformations of the four-velocity $X^{\alpha} = Y^{\alpha} = u^{\alpha} \rightarrow X^{\alpha} = Y^{\alpha} = u^{\alpha}+\Lambda^{\alpha}$ where we use the notation $\Lambda^{\alpha}=\Lambda_{,\beta}\:g^{\beta\alpha}$ for $\Lambda$ a local scalar, then on the local plane one the vectors that span this plane undergo a boost or a boost composed with a discrete full inversion or a boost composed with a discrete flip reflection leaving these vectors inside the original local plane. All these possible cases have been discussed in detail in manuscript \cite{A}. When we introduce gauge-like transformations of the four-velocity on the vectors that span the local plane two orthogonal to plane one, the vectors undergo a spatial rotation leaving them inside the original local plane two. This case has also been discussed in detail in manuscript \cite{A}. The metric tensor will remain a local gauge invariant under all these transformations as well as the four-velocity curl $u_{[\mu;\nu]}$ and the extremal field $\xi_{\mu\nu}$.

\section{Stress-energy tensor diagonalization tetrad}
\label{diagtetrad}

But these intermediate tetrad (\ref{Uw}-\ref{Ww}) is not the one that diagonalizes the stress-energy tensor. To this end, a new vector field can be defined through the expression,

\begin{eqnarray}
V_{(5)}^{\alpha} = V_{(4)}^{\alpha}\:(V_{(1)}^{\rho}\:u_{\rho}) -  V_{(1)}^{\alpha}\:(V_{(4)}^{\rho}\:u_{\rho}) \ . \label{v5}
\end{eqnarray}

Through the use of the antisymmetry of $\xi_{\mu\nu}$, the condition (\ref{scond}), the identity (\ref{i2}) and the definition of the vectors (\ref{V1}-\ref{V4}), it is simple to prove the following orthogonalities,

\begin{eqnarray}
u_{\mu}\:V_{(5)}^{\mu} = V_{(2)}^{\mu}\:g_{\mu\nu}\:V_{(5)}^{\nu} = V_{(3)}^{\mu}\:g_{\mu\nu}\:V_{(5)}^{\nu} = 0 \ . \label{ortho3}
\end{eqnarray}

Given that $u^{\mu}$, $V_{(2)}^{\mu}$, $V_{(3)}^{\mu}$ and $V_{(5)}^{\mu}$ are orthogonal, we can now proceed to see that these tetrad vectors covariantly and manifestly diagonalize the stress-energy tensor (\ref{SET}) at every spacetime point,

\begin{eqnarray}
u^{\alpha}\:T_{\alpha}^{\:\:\:\beta} &=& -(\rho + p)\:u^{\beta}
\label{EV1}\\
V_{(2)}^{\alpha}\:T_{\alpha}^{\:\:\:\beta} &=& p\:V_{(2)}^{\beta}
\label{EV2}\\
V_{(3)}^{\alpha}\:T_{\alpha}^{\:\:\:\beta} &=& p\:V_{(3)}^{\beta}
\label{EV3}\\
V_{(5)}^{\alpha}\:T_{\alpha}^{\:\:\:\beta} &=& p\:V_{(5)}^{\beta}\ .
\label{EV4}
\end{eqnarray}

Finally, we normalize this local tetrad,

\begin{eqnarray}
U^{\alpha} &=& u^{\alpha} \label{U}\\
V^{\alpha} &=& \xi^{\alpha\lambda}\:u_{\lambda} \:
/ \: (\:\sqrt{u_{\mu} \ \xi^{\mu\sigma} \
\xi_{\nu\sigma} \ u^{\nu}}\:) \label{V}\\
Z^{\alpha} &=& \ast \xi^{\alpha\lambda} \: u_{\lambda} \:
/ \: (\:\sqrt{u_{\mu}  \ast \xi^{\mu\sigma}
\ast \xi_{\nu\sigma}   u^{\nu}}\:)
\label{Z}\\
W^{\alpha} &=& \left( V_{(4)}^{\alpha}\:(V_{(1)}^{\rho}\:u_{\rho}) -  V_{(1)}^{\alpha}\:(V_{(4)}^{\rho}\:u_{\rho}) \right) / \: \sqrt{V_{(5)}^{\beta}\:V_{(5)_{\beta}} } \ ,
\label{W}
\end{eqnarray}

where, $V_{(5)}^{\beta}\:V_{(5)_{\beta}} = (V_{(4)}^{\beta}\:V_{(4)_{\beta}})\:(V_{(1)}^{\rho}\:u_{\rho})^{2} + (V_{(1)}^{\beta}\:V_{(1)_{\beta}})\:(V_{(4)}^{\rho}\:u_{\rho})^{2}$. It is obvious that $V^{\alpha}=\widehat{V}^{\alpha}$ and $Z^{\alpha}=\widehat{Z}^{\alpha}$ in equations (\ref{Vw}-\ref{Zw}).

\section{Equivalence of tetrad sets}
\label{equivalencetetrads}

We will show in this section how to prove in a straightforward fashion that the metric tensor written in terms of the tetrad sets (\ref{Uw}-\ref{Ww}) and (\ref{U}-\ref{W}) is exactly the same as expected. This proof will help us with the goals of the manuscript. Let us write instead of the vector $V_{(5)}^{\beta}$ an alternative vector that we will call $V_{(6)}^{\beta}$,

\begin{eqnarray}
V_{(6)\mu} = u_{\mu}\:(V_{(4)}^{\rho}\:u_{\rho}) + V_{(4)\mu} \ . \label{v5alt}
\end{eqnarray}

The orthogonalities in equation (\ref{ortho3}) will still be satisfied for the vector $V_{(6)}^{\mu}$, that is,
$u_{\mu}\:V_{(6)}^{\mu} = V_{(2)}^{\mu}\:g_{\mu\nu}\:V_{(6)}^{\nu} = V_{(3)}^{\mu}\:g_{\mu\nu}\:V_{(6)}^{\nu} = 0$. The metric tensor will be written in terms of this new tetrad version as,

\begin{equation}
g_{\alpha\beta} = -u_{\alpha}\:u_{\beta} + V_{\alpha}\:V_{\beta} +
Z_{\alpha}\:Z_{\beta} + {V_{(6)\alpha}\:V_{(6)\beta} \over (V_{(6)\sigma}\:V_{(6)}^{\sigma})}\ ,\label{NMT}
\end{equation}

These new tetrad vectors $(u_{\alpha}, V_{\alpha}, Z_{\alpha}, {V_{(6)\alpha} \over \sqrt{V_{(6)\sigma}\:V_{(6)}^{\sigma}}})$ will diagonalize the stress-energy tensor (\ref{SET}) at every spacetime point exactly as the tetrad set (\ref{U}-\ref{W}) given by $(u_{\alpha}, V_{\alpha}, Z_{\alpha}, {V_{(5)\alpha} \over \sqrt{V_{(5)\sigma}\:V_{(5)}^{\sigma}}})$ in equations (\ref{EV1}-\ref{EV4}). It is clear that we can write $u_{\mu}=(V_{(1)\mu}-V_{(4)\mu})/(Q/2)$ using the equations (\ref{V1}) and (\ref{V4}) with $X^{\alpha} = Y^{\alpha} = u^{\alpha}$. We find this result through the use of identity (\ref{i2}). We also find

\begin{equation}
V_{(6)\sigma}\:V_{(6)}^{\sigma}=  (V_{(4)}^{\rho}\:u_{\rho})^{2} + V_{(4)\sigma}\:V_{(4)}^{\sigma}\ . \label{v6norm}
\end{equation}

where $V_{(4)}^{\rho}$ has been gauged with $Y^{\alpha} = u^{\alpha}$. We simply present auxiliary results that will simplify the proof,

\begin{eqnarray}
V_{(4)\sigma}\:V_{(4)}^{\sigma} &=& V_{(3)\sigma}\:V_{(3)}^{\sigma} \label{4433} \\
V_{(3)\sigma}\:V_{(3)}^{\sigma} &=& (V_{(4)}^{\rho}\:u_{\rho})\:(-Q/2) \label{334u} \\
V_{(6)\sigma}\:V_{(6)}^{\sigma} &=& (V_{(4)}^{\rho}\:u_{\rho})^{2} + (-Q/2)\:(V_{(4)}^{\rho}\:u_{\rho}) \ . \label{66u4}
\end{eqnarray}

Making use of all these auxiliary results it is lengthy but rather straightforward to prove that the metric tensor expressed as in equation (\ref{NMT}) is exactly equal as the metric tensor expressed as in equation (\ref{MT}). This expected result is useful because we can then say that when we perform the transformation $u^{\alpha} \rightarrow u^{\alpha}+\Lambda^{\alpha}$ which works as a kind of ``gauge transformation'' as in electromagnetic theory with the electromagnetic potential, then the metric tensor in its expression (\ref{NMT}) will be as invariant as in expression (\ref{MT}), which already was known from manuscript \cite{A} to be invariant under this kind of transformation. We guarantee in this way that the not obvious invariance of the metric tensor in equation (\ref{NMT}) is so because we have proved the equality of both tetrad presentations, which by the way was expected even though not obvious as well.

\section{Vorticity stress-energy tensor invariant under gauge-like four-velocity transformations}
\label{gaugetransf}

We would like to study the geometry for a new choice in gauge vectors $X^{\alpha} = Y^{\alpha} = u^{\alpha} \rightarrow X^{\alpha} = Y^{\alpha} = u^{\alpha}+\Lambda^{\alpha}$ where we use the notation $\Lambda^{\alpha}=\Lambda_{,\beta}\:g^{\beta\alpha}$ for short. The metric tensor,

\begin{equation}
g_{\alpha\beta} = -U_{\alpha}\:U_{\beta} + V_{\alpha}\:V_{\beta} +
Z_{\alpha}\:Z_{\beta} + W_{\alpha}\:W_{\beta} = -\hat{U}_{\alpha}\:\hat{U}_{\beta} + \hat{V}_{\alpha}\:\hat{V}_{\beta} +
\hat{Z}_{\alpha}\:\hat{Z}_{\beta} + \hat{W}_{\alpha}\:\hat{W}_{\beta}\ ,\label{MTGT}
\end{equation}

will remain invariant under this new choice since it has already been proved in detail in reference \cite{A} that under a change of this kind, the metric tensor remains invariant, specially through equation or expression (\ref{MT}). The curl of the four-velocity $u_{[\mu;\nu]}$ will also remain invariant which is why the extremal field (\ref{vef}) and the scalar complexion (\ref{complexion}) will remain invariant as well. We resume our study and would like to know about the Einstein equations under this ``four-velocity gauge transformation''. The Einstein equations are given by,

\begin{eqnarray}
R_{\mu\nu} - \frac{1}{2}\:g_{\mu\nu}\:R &=& T_{\mu\nu}= (\rho + p)\:u_{\mu}\:u_{\nu} + p\:g_{\mu\nu}\ , \label{EFE}
\end{eqnarray}

as we notice from equation (\ref{SET}). Under the transformation $u^{\alpha} \rightarrow u^{\alpha}+\Lambda^{\alpha}$ we see that the right hand side will change because $u_{\mu}$ will change. The metric tensor remains invariant as already proved, therefore the left hand side of equation (\ref{EFE}) and the metric tensor on the second term of $T_{\mu\nu}$ will remain invariant. We will be able to write the transformed new stress-energy tensor as,

\begin{equation}
T^{new}_{\mu\nu}= (\rho + p)\:(u_{\mu}+\Lambda_{\mu})\:(u_{\nu}+\Lambda_{\nu}) + p\:g_{\mu\nu}\ .\label{SETMOD}
\end{equation}

Then the perfect fluid stress-energy tensor will not be invariant under gauge-like four-velocity transformations by itself. In order to make it invariant we would have to add to the right hand side of the Einstein equations in (\ref{EFE}) terms including heat flow currents $(\rho + p)\:(u_{\mu}\:h_{\nu} + u_{\nu}\:h_{\mu})$ where $u^{\mu}\:h_{\mu}=0$ is satisfied and terms with viscous stresses in the fluid $\tau_{\mu\nu}$ where $u^{\mu}\:\tau_{\mu\nu}=0$ is also satisfied, see references \cite{HS,CW,RW} and section \ref{app: I}. This kind of fluid will no longer be ideal, however we are focussed on the invariance geometrical properties of the Einstein equations under the gauge-like four-velocity transformations. In section \ref{app: I} a thorough discussion is provided on how an imperfect fluid stress-energy tensor can be invariant under gauge-like four-velocity local transformations $u^{\alpha} \rightarrow u^{\alpha}+\Lambda^{\alpha}$. It is concluded that specific transformations have to be simultaneously satisfied for the heat flow currents, the viscous stresses, the density and pressure for the right hand side of the Einstein equations for the imperfect fluid to be invariant under this kind of local transformation. In addition to this invariance analysis we would also like to know if we can build a symmetric stress-energy tensor invariant under these kind of transformations for the vorticity. There has been a discussion for many decades about the necessity to include in the right hand side of the Einstein equations a stress-energy tensor associated to vorticity. As a proof of these arguments we quote for example from reference \cite{GKB} ``It is taken for granted, in most expositions of fluid dynamics, that a deviatory stress cannot be generated by pure rotation, irrespective of the structure of the fluid simply on the grounds that there is no deformation of the fluid, however rigorous justification for this believe is elusive''. We also quote from \cite{TQH} ``Quantized vortices covering vastly different length scales have been found in both computational studies and experiments. They play a significant role in a remarkably diverse range of phenomena, including ultracold atomic Bose-Einstein condensates [13.31-13.33] superfluid liquid $^{3}He$ [13.34], type II low-temperature superconductors [13.35] and pulsars (rotating neutron stars) [13.36-13.37]''. We also add the stretching of vortex lines. As we can see, it is necessary first at the classical level to find a symmetric tensor that can play the role of vorticity stress-energy on the right hand side of the Einstein equations. Under all these arguments, see also references \cite{GKB,TQH,CD,GQW,MSS} and all the references therein, we would like to proceed in absolute analogy to the electromagnetic case presented in reference \cite{A} to introduce a proposal for this vorticity stress-energy tensor and to the study of its properties. Let us proceed to introduce the following symmetric tensor,

\begin{equation}
T^{vort}_{\mu\nu}=\xi_{\mu\lambda}\:\:\xi_{\nu}^{\:\:\:\lambda}
+ \ast \xi_{\mu\lambda}\:\ast \xi_{\nu}^{\:\:\:\lambda}\ .\label{TEMDR}
\end{equation}

With equations (\ref{scond}) and (\ref{i2}) it becomes trivial to prove that the tetrad sets (\ref{V1}-\ref{V4}) and (\ref{Uw}-\ref{Ww}) diagonalize locally and covariantly the stress-energy tensor (\ref{TEMDR}). We call this tensor a stress-energy tensor because it is built exactly as in Einstein-Maxwell spacetimes just replacing the electromagnetic four-potential by the four-velocity and using the inverse of the local duality rotation introduced in equation (\ref{vef}) $u_{[\mu;\nu]} = \cos\alpha \:\: \xi_{\mu\nu} + \sin\alpha \:\: \ast \xi_{\mu\nu}$ we get,

\begin{equation}
T^{vort}_{\mu\nu}=u_{[\mu;\lambda]}\:\:u_{[\nu;\rho]}\:g^{\rho\lambda}
+ \ast u_{[\mu;\lambda]}\:\ast u_{[\nu;\rho]}\:g^{\rho\lambda}=\xi_{\mu\lambda}\:\:\xi_{\nu}^{\:\:\:\lambda}
+ \ast \xi_{\mu\lambda}\:\ast \xi_{\nu}^{\:\:\:\lambda} \ .\label{TEMDRINV}
\end{equation}

Since we are discussing geometrical structures we leave for the moment possible constant units factors aside. It is also clear that this tensor is not the whole stress-energy tensor on the right hand side of the Einstein fluid equations. The whole tensor satisfying the conservation equations $T^{\mu\nu}_{\:\:\:\:\:;\nu}=0$ would include the perfect fluid terms plus heat flow plus viscous stress plus the vorticity stress-energy. Vectors (\ref{V1}-\ref{V2}) or the normalized (\ref{Uw}-\ref{Vw}) span the local plane one where all vectors are eigenvectors of the tensor (\ref{TEMDR}) with eigenvalue $Q/2=\xi_{\mu\lambda}\:\xi_{\mu}^{\:\:\:\lambda}/2$ which we assume to be $Q\neq0$. Vectors (\ref{V3}-\ref{V4}) or the normalized (\ref{Zw}-\ref{Ww}) span the local orthogonal plane two where all vectors are eigenvectors of the tensor (\ref{TEMDR}) with eigenvalue $-Q/2=-\xi_{\mu\lambda}\:\xi_{\mu}^{\:\:\:\lambda}/2$. Under the transformation $u^{\alpha} \rightarrow u^{\alpha}+\Lambda^{\alpha}$ the vectors that span the local plane one would undergo a hyperbolic rotation inside this plane while the vectors that span the local plane two would undergo a spatial rotation inside this second plane. Since the curl field $u_{[\mu;\nu]}$ is locally invariant under this transformation, the vorticity stress-energy (\ref{TEMDRINV}) is invariant as well. We assume for simplicity that the local plane one is spanned by a timelike and a spacelike tetrad vectors. The details for this proof have been worked out step by step in reference \cite{A} for the electromagnetic case which is identical in mathematical structure to our present vorticity case. The hidden principle that has been guiding us in our search for the vorticity stress-energy symmetric tensor is the principle of symmetry.

\section{Conclusions}
\label{conclusions}

The key to understand the concept that we are introducing in this manuscript is that there is for fluids with vorticity a sector of similar ideas as presented in the paper \cite{A} for the electromagnetic and gravitational fields in Einstein-Maxwell spacetimes. In Einstein-Maxwell spacetimes there is a non-trivial curl of the electromagnetic potential four-vector. There is manifest electromagnetic gauge invariance of the metric tensor as expressed in terms of tetrads of an analogous nature as to (\ref{V1}-\ref{V4}) or (\ref{Uw}-\ref{Ww}). The main difference between the Einstein-Maxwell spacetimes and the perfect fluid spacetimes is the stress-energy tensor. In Einstein-Maxwell spacetimes the stress-energy tensor is invariant under electromagnetic gauge transformations. In perfect fluid spacetimes the stress-energy tensor is not necessarily invariant under local gauge-like transformations of the four-velocity vectors. It would become necessary to introduce heat flows and viscous stresses to make it invariant. We proceeded in this direction in section \ref{app: I} where we introduced a new kind of local gauge transformation in the heat flux vector, the viscous stress-energy tensor, the density and pressure in order to make the whole imperfect fluid stress-energy tensor invariant. The point that we are addressing is that the Einstein imperfect fluid equations have a left hand side which is manifestly invariant under local four-velocity gauge-like transformations and therefore the right hand side, that is, the imperfect fluid stress-energy tensor must be also invariant when appropriate local transformations are implemented not only on the four-velocity but also on the heat flux, the viscous stress-energy, the density and pressure as well. In addition to this fact we presented a new vorticity stress-energy tensor built exactly as in the Einstein-Maxwell spacetimes given that we can replace the electromagnetic four potential for the fluid four-velocity, and save for multiplicative units constants, obtain similar tensor structures. This new vorticity stress-energy tensor is manifestly invariant under four-velocity local gauge-like transformations too and by itself. This was manifested mainly through the analysis in the section \ref{gaugetransf}. An application to neutron stars will be developed in section \ref{application} in order to show the simplifications brought about by these new tetrads. We quote from \cite{HK,HG} the Einstein Spencer Lecture in 1933, ``I am convinced that we can discover by means of purely mathematical constructions the concepts and the laws connecting them with each other, which furnish the key to the understanding of natural phenomena. Experience may suggest the appropriate mathematical concepts, but they most certainly cannot be deduced from it. Experience remains, of course, the sole criterion of physical utility of a mathematical construction. But the creative principle resides in mathematics. In a certain sense, therefore, I hold it true that pure thought can grasp reality, as the ancients dreamed''. We also quote from \cite{SW} ``Einstein made two heuristic and physically insightful steps. The first was to obtain the field equations in vacuum in a rather geometric fashion. The second step was obtaining the field equations in the presence of matter from the field equations in vacuum. This transition is an essential principle in physics, much as the principle of local gauge invariance in quantum field theory''.

\section{Appendix I: Stress-energy tensor symmetry}
\label{app: I}

Let us consider an imperfect fluid stress-energy tensor of the kind \cite{JO},

\begin{equation}
T^{imp}_{\mu\nu}= (\rho + p)\:u_{\mu}\:u_{\nu} + p\:g_{\mu\nu} + (q_{\mu}\:u_{\nu} + q_{\nu}\:u_{\mu}) + \tau_{\mu\nu}\ ,\label{SETIMP}
\end{equation}

where $q_{\mu}$ is the heat flux relative to $u_{\mu}$ and the viscous stress-energy tensor $\tau_{\mu\nu}$ is given by \cite{JO},

\begin{equation}
\tau_{\mu\nu}= -\eta\:\left(u_{\mu;\nu} + u_{\nu;\mu} + u_{\mu}\:u^{\alpha}\:u_{\nu;\alpha} + u_{\nu}\:u^{\alpha}\:u_{\mu;\alpha}\right) - (\zeta-\frac{2}{3}\:\eta)\:u^{\alpha}_{\:\:;\alpha}\:(g_{\mu\nu} + u_{\mu}\:u_{\nu}) \ .\label{SETIMPVISC}
\end{equation}

The parameter $\eta$ is the coefficient of shear viscosity and the parameter $\zeta$ is the coefficient of bulk viscosity. Under the four-velocity gauge-like transformation $u^{\alpha} \rightarrow u^{\alpha}+\Lambda^{\alpha}$ the stress-energy tensor (\ref{SETIMP}) will change and will not be invariant if this is the only transformation that we implement. Let us remember that the Einstein-Maxwell equations are given now by equation,

\begin{eqnarray}
R_{\mu\nu} - \frac{1}{2}\:g_{\mu\nu}\:R &=& T^{imp}_{\mu\nu} \ . \label{EFEIMP}
\end{eqnarray}

The left hand side of equations (\ref{EFEIMP}) will remain invariant since it has been proven already in sections \ref{diagtetrad}-\ref{equivalencetetrads} and reference \cite{A} that the metric tensor is manifestly invariant under $u^{\alpha} \rightarrow u^{\alpha}+\Lambda^{\alpha}$. The covariant derivatives, that is, the Christoffel symbols will also be invariant for the same reason. Therefore by this symmetry argument we observe that the right hand side of equations (\ref{EFEIMP}) has also to be invariant on its own. When we carry out the transformations $u^{\alpha} \rightarrow u^{\alpha}+\Lambda^{\alpha}$ on the right hand side we notice that the tensor $T^{imp}_{\mu\nu}$ is not invariant only under this kind of transformation. In order to make it invariant we introduce the following additional local transformations,

\begin{eqnarray}
\rho &\rightarrow& \rho + \widetilde{\rho} \label{rhotransf}\\
p &\rightarrow& p + \widetilde{p} \label{ptransf}\\
q^{\mu} &\rightarrow& q^{\mu} + \widetilde{q}^{\mu} \label{heattransf}\\
\tau^{\mu\nu}(u) &\rightarrow& \tau^{\mu\nu}(u+\Lambda) + \widetilde{\tau}^{\mu\nu} \ . \label{viscoustransf}
\end{eqnarray}

By $\tau^{\mu\nu}(u)$ we mean exactly the expression (\ref{SETIMPVISC}) while by $\tau^{\mu\nu}(u+\Lambda)$ we mean expression (\ref{SETIMPVISC}) under the transformation $u^{\alpha} \rightarrow u^{\alpha}+\Lambda^{\alpha}$. This way we introduce notation that will shorten the explicit writing of long expressions with many terms although straightforwardly clear in its content. We also know that there is a state equation $p(\rho)$ and there is also another one $\widetilde{p}(\widetilde{\rho})$. There is then only one local scalar independent variable that we might consider to be $\widetilde{\rho}$. When we impose invariance of the stress-energy tensor $T^{imp}_{\mu\nu}$ we will obtain ten equations for the fifteen variables $\widetilde{\rho}$, $\widetilde{q}^{\mu}$ and $\widetilde{\tau}^{\mu\nu}$. In addition we also know that $q^{\alpha}\:u_{\alpha}=0$ and $\tau^{\mu\nu}\:u_{\nu}=0$ which are five more equations. When we extend these last two equations to the case under $u^{\alpha} \rightarrow u^{\alpha}+\Lambda^{\alpha}$ we find $(q^{\mu} + \widetilde{q}^{\mu})\:(u_{\mu}+\Lambda_{\mu})=0$ and $(\tau^{\mu\nu}(u+\Lambda) + \widetilde{\tau}^{\mu\nu})\:(u_{\mu}+\Lambda_{\mu})=0$. From these last two ``transverse'' equations plus the original ones $q^{\alpha}\:u_{\alpha}=0$ and $\tau^{\mu\nu}\:u_{\nu}=0$ we obtain the following,

\begin{eqnarray}
q^{\mu}\:\Lambda_{\mu} + \widetilde{q}^{\mu}\:(u_{\mu}+\Lambda_{\mu})&=&0 \label{conditionsheat}\\
\tau^{\mu\nu}(u)\:\Lambda_{\nu} + \tau^{\mu\nu}(\Lambda)\:(u_{\nu}+\Lambda_{\nu}) + \widetilde{\tau}^{\mu\nu}\:(u_{\nu}+\Lambda_{\nu})&=&0 \ . \label{conditionsvisc}
\end{eqnarray}

By $\tau^{\mu\nu}(\Lambda)$ we mean all the terms of the style $\Lambda_{\mu}\:u_{\nu;\alpha}\:u^{\alpha}+\Lambda_{\nu}\:u_{\mu;\alpha}\:u^{\alpha}$ or $u_{\mu}\:\Lambda_{\nu;\alpha}\:\Lambda^{\alpha}+u_{\nu}\:\Lambda_{\mu;\alpha}\:\Lambda^{\alpha}$ or $\Lambda_{\mu}\:\Lambda_{\nu;\alpha}\:\Lambda^{\alpha}+\Lambda_{\nu}\:\Lambda_{\mu;\alpha}\:\Lambda^{\alpha}$ just to show a few examples. The tensor $\tau^{\mu\nu}(u)$ is just (\ref{SETIMPVISC}) and we can write $\tau^{\mu\nu}(u+\Lambda)=\tau^{\mu\nu}(u)+\tau^{\mu\nu}(\Lambda)$. We proceed then to implement in expression (\ref{SETIMP}) the simultaneous transformations $u^{\alpha} \rightarrow u^{\alpha}+\Lambda^{\alpha}$ plus transformations (\ref{rhotransf}-\ref{viscoustransf}). We find,

\begin{eqnarray}
T^{imp\:\mu\nu} \rightarrow T^{imp\:\mu\nu} &+& (\rho + p)\:(u^{\mu}\:\Lambda^{\nu} + u^{\nu}\:\Lambda^{\mu} + \Lambda^{\mu}\:\Lambda^{\nu}) \nonumber \\ &+& (\widetilde{\rho} + \widetilde{p})\:(u^{\mu}+\Lambda^{\mu})\:(u^{\nu}+\Lambda^{\nu}) + \widetilde{p}\:g^{\mu\nu}
+ \Lambda^{\mu}\:q^{\nu} + \Lambda^{\nu}\:q^{\mu} + \nonumber \\ &+& \widetilde{q}^{\mu}\:(u^{\nu}+\Lambda^{\nu}) + \widetilde{q}^{\nu}\:(u^{\mu}+\Lambda^{\mu}) + \tau^{\mu\nu}(\Lambda) + \widetilde{\tau}^{\mu\nu} \ . \label{eqTimp}
\end{eqnarray}

Next, we impose invariance,

\begin{eqnarray}
&&(\rho + p)\:(u^{\mu}\:\Lambda^{\nu} + u^{\nu}\:\Lambda^{\mu} + \Lambda^{\mu}\:\Lambda^{\nu}) \nonumber \\ &+& (\widetilde{\rho} + \widetilde{p})\:(u^{\mu}+\Lambda^{\mu})\:(u^{\nu}+\Lambda^{\nu}) + \widetilde{p}\:g^{\mu\nu} + \Lambda^{\mu}\:q^{\nu} + \Lambda^{\nu}\:q^{\mu} \nonumber \\ &+& \widetilde{q}^{\mu}\:(u^{\nu}+\Lambda^{\nu}) + \widetilde{q}^{\nu}\:(u^{\mu}+\Lambda^{\mu}) + \tau^{\mu\nu}(\Lambda) + \widetilde{\tau}^{\mu\nu} = 0 \ . \label{eqTimpinv}
\end{eqnarray}

If we contract equation (\ref{eqTimpinv}) with $(u_{\nu}+\Lambda_{\nu})$ and make use of equations (\ref{conditionsheat}-\ref{conditionsvisc}) we find the object $\widetilde{q}^{\mu}$,
\begin{eqnarray}
(u^{\nu}+\Lambda^{\nu})\:(u_{\nu}+\Lambda_{\nu})\:\widetilde{q}^{\mu} &=& \{(q^{\alpha}\:\Lambda_{\alpha})\:(u^{\mu}+\Lambda^{\mu}) + \tau^{\mu\alpha}(u)\:\Lambda_{\alpha} \nonumber \\ &-& [(\rho + p)\:(u^{\mu}\:\Lambda^{\alpha} + u^{\alpha}\:\Lambda^{\mu} + \Lambda^{\mu}\:\Lambda^{\alpha}) + \Lambda^{\mu}\:q^{\alpha} + \Lambda^{\alpha}\:q^{\mu} \nonumber \\ &+& (\widetilde{\rho} + \widetilde{p})\:(u^{\mu}+\Lambda^{\mu})\:(u^{\alpha}+\Lambda^{\alpha}) + \widetilde{p}\:g^{\mu\alpha}]\:(u_{\alpha}+\Lambda_{\alpha})\} \ . \label{qtilde}
\end{eqnarray}

If we contract one more time with $(u_{\mu}+\Lambda_{\mu})$ and use again the condition (\ref{conditionsheat}) we can obtain the equation corresponding to $\widetilde{\rho}$ and $\widetilde{p}$,

\begin{eqnarray}
(u^{\nu}+\Lambda^{\nu})\:(u_{\nu}+\Lambda_{\nu})\:(-q^{\alpha}\:\Lambda_{\alpha}) &=& \{(q^{\alpha}\:\Lambda_{\alpha})\:(u^{\mu}+\Lambda^{\mu})\:(u_{\mu}+\Lambda_{\mu}) + \tau^{\mu\alpha}(u)\:\Lambda_{\alpha}\:(u_{\mu}+\Lambda_{\mu}) \nonumber \\ &-& [(\rho + p)\:(u^{\mu}\:\Lambda^{\alpha} + u^{\alpha}\:\Lambda^{\mu} + \Lambda^{\mu}\:\Lambda^{\alpha}) + \Lambda^{\mu}\:q^{\alpha} + \Lambda^{\alpha}\:q^{\mu} \nonumber \\ &+& (\widetilde{\rho} + \widetilde{p})\:(u^{\mu}+\Lambda^{\mu})\:(u^{\alpha}+\Lambda^{\alpha}) + \widetilde{p}\:g^{\mu\alpha}]\: . \nonumber \\ &.& (u_{\alpha}+\Lambda_{\alpha})\:(u_{\mu}+\Lambda_{\mu})\} \ . \label{qtilde2}
\end{eqnarray}

We first find $\widetilde{p}(\widetilde{\rho})$ from equation (\ref{qtilde2}),

\begin{eqnarray}
&&[(\widetilde{\rho} + \widetilde{p})\:(u^{\mu}+\Lambda^{\mu})\:(u^{\alpha}+\Lambda^{\alpha}) + \widetilde{p}\:g^{\mu\alpha}]\:(u_{\alpha}+\Lambda_{\alpha})\:(u_{\mu}+\Lambda_{\mu}) = \nonumber \\ &&\{(q^{\alpha}\:\Lambda_{\alpha})\:(u^{\mu}+\Lambda^{\mu})\:(u_{\mu}+\Lambda_{\mu}) + \tau^{\mu\alpha}(u)\:\Lambda_{\alpha}\:(u_{\mu}+\Lambda_{\mu}) \nonumber \\ &-& [(\rho + p)\:(u^{\mu}\:\Lambda^{\alpha} + u^{\alpha}\:\Lambda^{\mu} + \Lambda^{\mu}\:\Lambda^{\alpha}) + \Lambda^{\mu}\:q^{\alpha} + \Lambda^{\alpha}\:q^{\mu}]\: . \nonumber \\ &.& (u_{\alpha}+\Lambda_{\alpha})\:(u_{\mu}+\Lambda_{\mu})\} - (u^{\nu}+\Lambda^{\nu})\:(u_{\nu}+\Lambda_{\nu})\:(-q^{\alpha}\:\Lambda_{\alpha}) \ , \label{ptilde3}
\end{eqnarray}

and as we are supposed to know this equation of state $\widetilde{p}(\widetilde{\rho})$ from the outset in an independent fashion, we also get to find $\widetilde{\rho}$ just by equating these two independent expressions of $\widetilde{p}(\widetilde{\rho})$ and then replacing both $\widetilde{p}(\widetilde{\rho})$ and $\widetilde{\rho}$ back in equation (\ref{qtilde}). Once we found $\widetilde{q}^{\mu}$ in equation (\ref{qtilde}) we replace it along with $\widetilde{p}(\widetilde{\rho})$ and $\widetilde{\rho}$ in equation (\ref{eqTimpinv}) and find $\widetilde{\tau}^{\mu\nu}$. Let us review the rational behind all of these procedures. We start with the imperfect stress-energy tensor (\ref{SETIMP}). Then we implement simultaneous transformations $u^{\alpha} \rightarrow u^{\alpha}+\Lambda^{\alpha}$ plus transformations (\ref{rhotransf}-\ref{viscoustransf}). The original (\ref{SETIMP}) plus the four-velocity gauge-like transformation scalar $\Lambda$ or the gradient $\Lambda_{\mu}$ are all known as are the equations of state $p(\rho)$ and $\widetilde{p}(\widetilde{\rho})$.
But then we also get to know $\widetilde{p}(p(\rho),\rho,\widetilde{\rho})$ from the outset. In fact from equation (\ref{ptilde3}) we observe that $\widetilde{p}$ is $\widetilde{p}(p(\rho),\rho,\widetilde{\rho})$. These are the two $\widetilde{p}$ that we equate in order to obtain $\widetilde{\rho}$ in terms of the objects given at the outset. Next we impose invariance through equations (\ref{eqTimpinv}) under all of these local transformations. Finally and using the conditions (\ref{conditionsheat}-\ref{conditionsvisc}) we obtain by algebraic work both $\widetilde{q}^{\mu}$ and $\widetilde{\tau}^{\mu\nu}$ plus $\widetilde{p}(\widetilde{\rho})$ and $\widetilde{\rho}$. We used ten equations or invariance conditions through equations (\ref{eqTimpinv}) plus the five conditions (\ref{conditionsheat}-\ref{conditionsvisc}) and found fifteen local objects $\widetilde{q}^{\mu}$, $\widetilde{\tau}^{\mu\nu}$ and $\widetilde{p}(\widetilde{\rho})$ or $\widetilde{\rho}$. This method ensures the gauge-like invariance in the right hand side of the system (\ref{EFEIMP}) knowing a priori that the left hand side of the system (\ref{EFEIMP}) is explicitly and manifestly invariant under this local group of transformations since this latter claim has been proven in sections \ref{diagtetrad}-\ref{equivalencetetrads} and reference \cite{A}. In the end we are also able to obtain different tetrad states of spacetime for the same system of equations (\ref{EFEIMP}) which is an additional result in this manuscript so that a posteriori we can focus on the proper and improper local tetrad transformations on plane one, such as boosts, inversions and flips.

\section{Appendix II: Application to neutron stars}
\label{application}

In this section we will study the components of the Ricci tensor when the imperfect fluid terms in equation (\ref{SETIMP}) and equation (\ref{SETIMPVISC}) can be neglected. We will keep the perfect fluid terms plus the vorticity terms. We do not mean that the local four-velocity gauge-like symmetry that we have been studying in this manuscript ceases to exist, rather we consider that even though the symmetry still exists, for practical purposes the stress-energy tensor imperfect fluid terms can be neglected with respect to the perfect fluid stress-energy terms (\ref{SET}) for example. We will focus on this kind of simplifications. This situation might happen in neutron stars and there is an important amount of literature, we just cite some papers and reviews where more references can be found \cite{F8,NS,BC,RS,NC,AC2,FI,AWV,DL,291}. In some of this literature there are even comparisons between the use of the perfect fluid only and perfect fluid plus imperfect fluid corrections.  We implement this new technique using only covariant and local manipulations of an algebraic nature, which will not add more substantial computational time and nonetheless bring about simplification in further applications like in spacetime dynamical evolution. The stress-energy tensor under all of the above considerations would be given by,

\begin{equation}
T_{\mu\nu} = (\rho + p)\:u_{\mu}\:u_{\nu} + p\:g_{\mu\nu} + T^{vort}_{\mu\nu}\ ,\label{SETVORT}
\end{equation}

where the vorticity tensor is provided in equation (\ref{TEMDR}). If we call $Q^{vort}=\xi_{\mu\lambda}\:\:\xi^{\mu\lambda}$ and we keep in mind that the metric tensor is given by equation (\ref{MTGT}) we can write the tensor (\ref{SETVORT}) as,

\begin{equation}
T_{\mu\nu} = (\rho + p)\:u_{\mu}\:u_{\nu} + p\:g_{\mu\nu} + \frac{Q^{vort}}{2}\:[-\hat{U}_{\mu}\:\hat{U}_{\nu} + \hat{V}_{\mu}\:\hat{V}_{\nu} -
\hat{Z}_{\mu}\:\hat{Z}_{\nu} - \hat{W}_{\mu}\:\hat{W}_{\nu}] \ .\label{SETVORTEXP}
\end{equation}

We can also verify using the identity equations (\ref{scond}) and (\ref{i2}) that,

\begin{eqnarray}
\hat{U}^{\alpha}\:T_{\alpha}^{\:\:\:\beta} &=& (\rho + p)\:(\hat{U}^{\alpha}\:u_{\alpha})\:u^{\beta} + [p+\frac{Q^{vort}}{2}]\:\hat{U}^{\beta}
\label{PVSETU}\\
\hat{V}^{\alpha}\:T_{\alpha}^{\:\:\:\beta} &=& [p+\frac{Q^{vort}}{2}]\:\hat{V}^{\beta}
\label{PVSETV}\\
\hat{Z}^{\alpha}\:T_{\alpha}^{\:\:\:\beta} &=& [p+\frac{Q^{vort}}{2}]\:\hat{Z}^{\beta}
\label{PVSETZ}\\
\hat{W}^{\alpha}\:T_{\alpha}^{\:\:\:\beta} &=&  (\rho + p)\:(\hat{W}^{\alpha}\:u_{\alpha})\:u^{\beta} + [p-\frac{Q^{vort}}{2}]\:\hat{W}^{\beta}\ .
\label{PVSETW}
\end{eqnarray}

We can also write $u^{\beta}=-(\hat{U}^{\alpha}\:u_{\alpha})\:\hat{U}^{\beta}+(\hat{W}^{\alpha}\:u_{\alpha})\:\hat{W}^{\beta}$. We already know that $\hat{V}^{\alpha}\:u_{\alpha}=0$ and $\hat{Z}^{\alpha}\:u_{\alpha}=0$. Using this last result and the equations (\ref{PVSETU}-\ref{PVSETW}) we conclude that only five components of the stress-energy tensor (\ref{SETVORT}) or the equivalent (\ref{SETVORTEXP}) are not zero using this new tetrad and this is an important source of simplification. The tetrad that we have found (\ref{Uw}-\ref{Ww}) will be used in order to find the components of the left hand side of the Einstein equations and see if there is a possible simplification precisely on the left hand side with independence of our stress-energy tensor method of diagonalization on the right hand side of the Einstein equations. Let us consider the following equation that can be found in the literature \cite{CW,WE},

\begin{eqnarray}
v_{\mu;\nu;\rho}-v_{\mu;\rho;\nu} = -R^{\sigma}_{\:\:\:\mu\nu\rho}\:v_{\sigma} \ . \label{dcdanti}
\end{eqnarray}

This equation (\ref{dcdanti}) is valid in general for any vector field $v_{\sigma}$. Let us see an example of simplification and use the previous algorithm starting with equation (\ref{dcdanti}) applied to the vector (\ref{Vw}) after contracting this equation on both sides with $g^{\mu\rho}$,

\begin{eqnarray}
\hat{V}^{\mu}_{\:\:\:;\nu;\mu}-\hat{V}^{\mu}_{\:\:\:;\mu;\nu} = -R^{\sigma}_{\:\:\:\nu}\:\hat{V}_{\sigma} \ . \label{dcdantikillVR}
\end{eqnarray}

Next, let us contract with vector (\ref{Zw}),

\begin{eqnarray}
\hat{Z}^{\nu}\:\hat{V}^{\mu}_{\:\:\:;\nu;\mu}-\hat{Z}^{\nu}\:\hat{V}^{\mu}_{\:\:\:;\mu;\nu} = -\hat{Z}^{\nu}\:R^{\sigma}_{\:\:\:\nu}\:\hat{V}_{\sigma} = -\hat{Z}^{\nu}\:(T_{\:\:\:\nu}^{\sigma}-\frac{1}{2}\:\delta^{\sigma}_{\:\:\:\nu}\:T^{\mu}_{\:\:\:\mu})\:\hat{V}_{\sigma} = 0 \ , \label{EEVZ}
\end{eqnarray}

since $\hat{Z}^{\alpha}\:T_{\alpha}^{\:\:\:\beta} = [p+\frac{Q^{vort}}{2}]\:\hat{Z}^{\beta}$ according to equation (\ref{PVSETZ}) and also using the orthogonality $\hat{Z}^{\sigma}\:\hat{V}_{\sigma} = 0$. Equation (\ref{EEVZ}) is another source of simplification. This idea can be repeated with other components as well.


\end{document}